\documentclass[journal]{IEEEtran}
\IEEEoverridecommandlockouts
\usepackage{cite}
\usepackage{amsmath,amssymb,amsfonts}
\usepackage{algorithmic}
\usepackage[]{subfigure}
\usepackage{multirow}
\usepackage{graphicx}
\usepackage{textcomp}     
\usepackage{makecell}
\usepackage{eurosym}
\usepackage{tikz}
\usetikzlibrary{arrows,shapes,positioning,shadows,trees}
\usetikzlibrary{shapes.geometric, arrows}        
\usepackage{forest}
\usetikzlibrary{mindmap,trees}             
\usepackage{xcolor}
\usepackage[ruled,vlined]{algorithm2e}
\usepackage{float}
\usepackage{array}
\usepackage{multirow,multicol}
\usepackage{eurosym}
\usepackage{siunitx}
\usepackage{booktabs}
\usepackage{caption}
\usepackage{tikz}
\newbox\tempbox%
\def\BibTeX{{\rm B\kern-.05em{\sc i\kern-.025em b}\kern-.08em
    T\kern-.1667em\lower.7ex\hbox{E}\kern-.125emX}}
    
\begin{document}

\title{Flexicurity: Unlocking the True Potential of Power Market Players in the European Smart Grid Era (White paper)
\thanks{I. Avramidis is with the Luxembourg Institute of Science and Technology, Belvaux, Luxembourg, and with KU Leuven, Leuven, Belgium (jasonvra@yahoo.com) G. Takis-Defteraios is with Statkraft A.S., London, United Kingdom.}
}

\author{\IEEEauthorblockN{Iason-Iraklis Avramidis and Gerasimos Takis-Defteraios}}
\maketitle

\begin{abstract}
As the "smart grid" paradigm becomes more prevalent, fundamental techno-economic challenges prominently arise. The variability of renewables may require conventional generators to remain active and operate inefficiently. The grid's inertia grows weaker, jeopardizing its stability. Network-related problems become more prevalent, threatening the grid’s security. Finally, the industry experiences price cannibalization to extensive degrees, with frequent negative prices and with renewable energy producers downregulating their production in order to avoid severe penalties. To this day, no previous work has provided a holistic analysis on how each energy asset could contribute to grid management within the power market setting. This paper attempts to fill this knowledge gap by investigating the ``flexicurity" potential of energy assets, in the context of smart grids and the so-called "2050 strategy". More light is cast on their true value, thus better guiding network operators and market players. To the above, we categorize all flexibility and security capabilities of each energy asset, qualitatively grade their ``performance". We secondly introduce the concept of ``flexicurity" and discuss a gradual shift towards an alternative flexicurity-based market operation. Considering desired state of the European power grid by 2050, we finally provide a flexicurity-based overview of possible asset integration routes.
\end{abstract}


\section{Introduction}

``Sustainable energy", ``smart grid", ``2050 strategy". Terms that even the scientifically unsophisticated reader has arguably become acquainted with. While they were once nothing sort of ``fantasies", echoed by only a handful of pioneers in the field, they have undeniably become the main ``agents" of innovation in the power systems community. In Europe (and beyond), massively ambitious targets have been set for renewable energy sources (RES), the end-goal being that they will be able to cover more than 20\% and 27\% of the energy consumption in the EU by 2020 and 2030, respectively. In fact, a substantial number of countries has even surpassed these targets, with ``hot" debates currently taking place regarding the potential of reaching 50\%-100\% targets by 2050, thus achieving the vision expressed over 45 years ago by Bent Sorensen \cite{Sorensen}.

So, is this revolutionary transformation that electricity systems are currently experiencing a fairytale-like development? Not exactly. The extensive decarbonization efforts and the deregulation paradigm have undoubtedly given rise to a rather colorful assortment of technoeconomic issues:  the variability and limited predictability and controllability of RES increased the need for balancing services, requiring hundreds of  conventional generators to remain active and operate inefficiently (partial-loading, frequent start-up and shut-down cycles). The introduction of inverter-based generation weakened the power grid’s inertia, leading to rapid frequency changes, and threatening much more openly the grid’s stability. Network-related issues, such as line congestions and voltage violations became more prevalent, throwing into question the adequacy of the grid’s security. Even the industry was not exempt from the effects of the previously mentioned transformation. The extensive degrees of frequent price cannibalization (with regularly occurring negative electricity prices) has brought about drops in profit, while RES owners have become keen to heavily down-regulate their production in order to avoid severe financial losses. 

There are myriads of works that have investigated the presence and impact of different energy assets within power markets and the role that they can play in managing and securing power systems. However, few, if any, have actually provided a comprehensive analysis of the existing situation from a holistic perspective. This paper makes the very first attempt to fill this knowledge gap by proposing a new concept for analyzing the benefit of each energy asset in terms of their ``flexicurity" potential, i.e., common evaluation of flexibility and security, procured as a single, unified service.  The idea is to support, in more market-efficient manner, the long-term goal of driving the power system closer to the 50\%-100\% penetration level of RES by 2050. The above involve the analysis of each energy asset terms such as type, size, reliability, sustainability, financial characteristics, flexibility support and contribution to system security. The ``flexicurity" evaluation sheds new light on the role and value of each asset and serves as a guiding force for network operators, market players, investors and decisions makers.

Considering the above goals, this work justifies its necessity through the 3 following points: 

\begin{itemize}
    \item It provides a comprehensive overview of the state of the European power grid and the status of all energy assets in the smart grid context.
    \item It raises awareness of the key weaknesses of current practices in power markets, and proposes, for the very first time, the unifying concept of ``flexicurity" as a superior alternative to evaluate the technical and financial viability of each asset.
    \item Under the umbrella of flexicurity, it offers insights on the ideal rollout and integration of new energy technologies, as it pertains to the 2050 strategy.
\end{itemize}

The reader should be wary before continuing. This is not a pure research paper. No fancy formulations will be presented, nor will any grandiose claims of methodological contributions be made. These few pages group together and outline in a structured manner the concerns of several members of academia and industry; for better or worse, all concerned parties simply happen to be indirectly represented by the authors. Our aim is to raise awareness of fundamental issues that have not been payed proper attention thus far. We hope to inspire future researchers to acknowledge these issues when performing technical research; this way we will have perhaps added our own little steppingstone towards bridging the academia-industry gap. 

The rest of the paper is organized as follows. Section 2 gives a comprehensive overview of the current state of the European power grid; key issues are illustrated and analyzed. Section 3 presents the concept of ``flexicurity" and its advantages over current practices; given that the concept does not really exist, the analysis is predominantly qualitative. Section 4 discusses the subject of how the power grid can move toward the envisioned 2050 state, within the context of flexicurity. Section 5 recapitulates the main points of the paper and concludes by highlighting relevant research directions that the community should follow in the future

\section{The state of the European power grid}

Let us start off by looking at the current state (as of 2021) of the European power grid, i.e., the massive ``creature" that is being managed by ENTSO-E, dozens of TSOs and hundreds of DSOs. Fig. 1 will be our guide at this stage. So, who are the dominant players at the moment? Well, coal/gas, nuclear and hydro power plants (which we will call conventional for simplicity) maintain a very strong presence in the grid ($\sim$65\%), the distribution of course not being uniform across countries. Industry experience has shown that operators consistently make overly conservative decisions with respect to the reserves procured. In an effort to over-secure the system, these long-term decisions restrict conventional plants in terms of flexibility, ``force" them to withhold capacity, and effectively stop them from participating in many market processes. Even though RES, i.e., solar, wind, geothermal, and sea current energy, have a strong presence in theory ($\sim$30\%), this is but a half-truth; curtailment orders, down-regulation decisions and under-prediction strategies during bid submissions actually lead to utilization that is representative of lower penetration levels ($<$20\%). 

\begin{figure}[t!] 
	\centering
	\scalebox{0.5}{\includegraphics[]{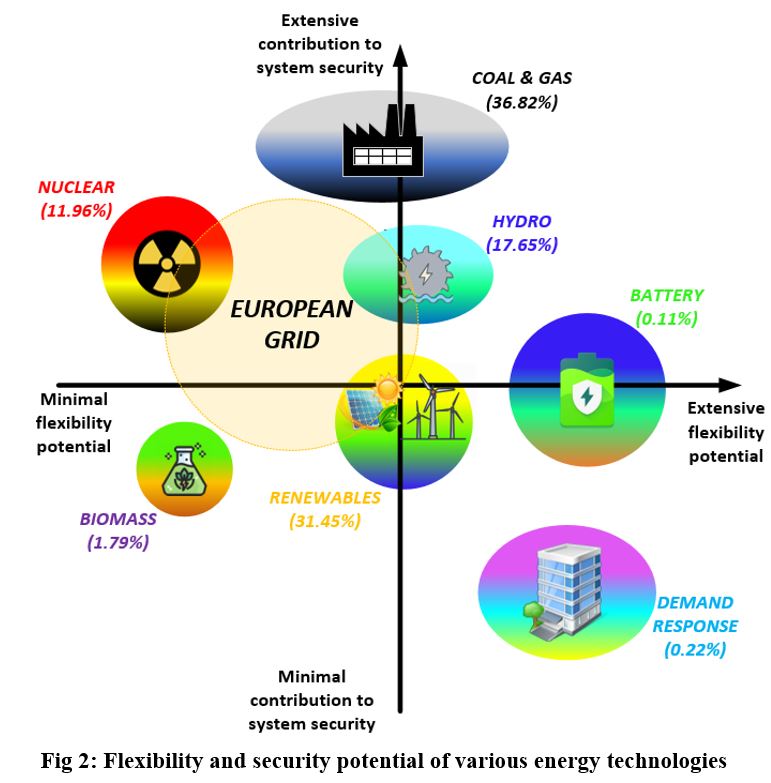}}
	\caption{Overview of the current (2021) European power grid: flexibility and security potential of individual energy assets. Note that the potential is with respect to what the asset could contribute under full deployment, not with respect to what it contributes under its current penetration level.}
	\label{Techniques}
\end{figure}

The ``game-changers" that were promised to drastically increase the grid’s flexibility and reduce the stress of over-securing the system, i.e., battery storage system and demand response participants, are practically non-existent, with an active penetration of less than 0.4\%. Demand response sees limited application, mostly through aggregators, voluntary actions by customers, and dynamic pricing schemes of varying degrees of success. No coordinated effort for their integration has truly been made. The industry has been very hesitant in the adoption of large-scale batteries, mostly due to the currently high costs and the unclear and restrictive legal situation across the continent. The few successful applications concern self-regulation (virtual power plants), some stand-alone market participation, and within distribution networks, where they find various applications \cite{Koller}. Still, a far cry away from what the academic community may think is the case.  

People like to talk about a European interconnected system, but this has yet to be implemented to a drastic degree. Keep in mind that each country is still largely responsible for defining its requirements with respect to security and flexibility guidelines, as well as for designing its own market-clearing rules and regulations. To make matters worse, the overlying market-clearing mechanism is not yet 100\% uniform across the continent. For example, up until only a few months ago, Greece was still using unit commitment on a national level, despite an adaptation directive (to the envisioned European standards) having been issued years ago. Furthermore, energy traders and large market players have actually further exasperated market inefficiencies and substantially expanded the role of reserve markets, simply because they have yet to adapt or to fully understand how the new market framework works, resulting in grid-unfriendly decisions.

In conclusion, if we combine all of the above, plus the testimonies of real-life market players (shown below in quotations), what can we conclude? Five major points consistently turn up, highlighting  the following fundamental concerns:

\begin{enumerate}
    \item The power grid is overly secured, with many market players being ``inefficiently utilized" and ``locked out of many market processes".
    \item Contrary to popular belief, the grid’s flexibility levels are not in par with those of security, hence the grid’s inability to ``react as efficiently as it could" and the operators’ ``insistence on preparing for doomsday events that are simply infeasible".
    \item RES players are not as prevalent as they could be; they consistently fall victims to under-utilization directives and suffer negative impacts from ``flawed market processes".
    \item Technologies that could revolutionize the system under substantial penetrations (batteries and demand response) are characterized by limited ``willingness to adopt", ``legal assistance", and ``coordination efforts".
    \item Academia ``consistently ignores fundamental aspects of modern power systems operation" and focuses on methodologies and applications that are ``not foreseen to be applicable in the near future" or are ``too detached from reality".
\end{enumerate}

\section{``Flexicurity": A necessary step for taking power markets to the next level}

The question now becomes what we can do to combat the issues above. Obviously, we cannot re-create the power market anew. We can however adapt it to fit its true needs. Some recent pioneering works have discussed the potential of shifting to new market paradigms, such as introducing real-time scarcity prices \cite{Papavasiliou}, or sharpening the focus to more rigorously evaluate the grid’s reactive capabilities as we move to RES-heavy situations \cite{Capitanecu}. These works are certainly good strides towards the right direction. However, we argue that the good aspects that they introduce can in fact be combined to form a single entity. This is where the concept of ``flexicurity" comes into play, at which point we ask ourselves: ``how would this new product actually look like and how would it be dealt with"?

\begin{figure}[b!] 
	\centering
	\scalebox{0.4}{\includegraphics[]{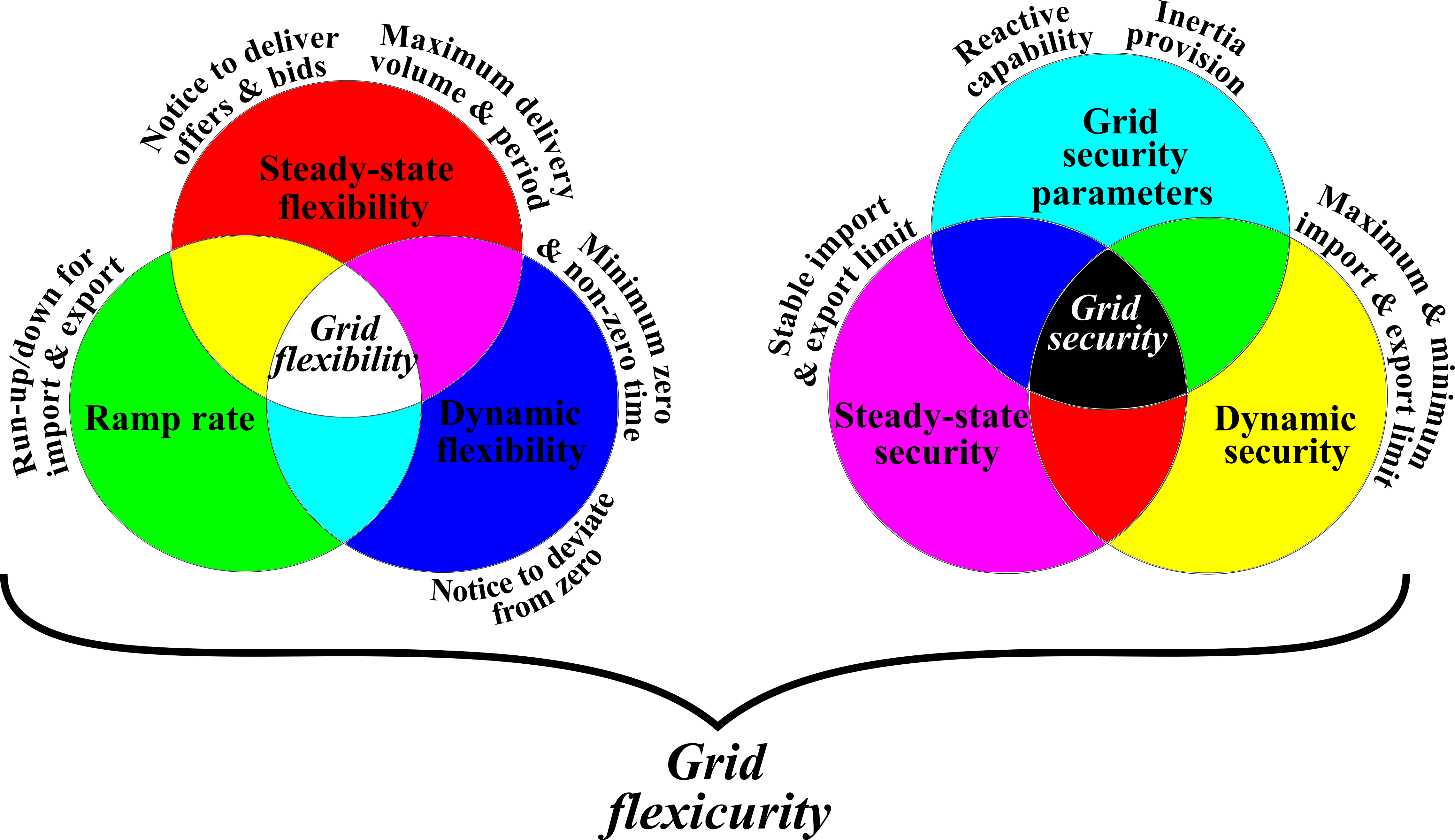}}
	\caption{Flexicurity concept visualization: breakdown between flexibility and security and their specific grid-driven, market-based services}
	\label{Techniques1}
\end{figure}

We refer the reader to Fig. 2, which presents all currently tradeable products with respect to grid flexibility and grid security. Evidently, even when dealing with a single element of power system preparation (e.g., sufficient steady-state security), system operators have to consider several aspects before making a procurement decision, especially when several market players are involved. These concerns are further magnified and show their ``ugliest" face when operators are faced with the monumental task of ensuring a secure and flexible grid from every viewpoint. This requires dozens of separate markets, the balancing of hundreds of market players and the careful evaluation of thousands of technical characteristics. Sadly, inefficient and sub-optimal decisions are the nigh-exclusive norm in such cases.

We now turn the reader’s attention to Tables 1, 2 below, which shows a first crude way of looking at the potential of each energy aspect. Note that the ``grading scheme" below is purely empirical, stemming from experience accumulated from both industry and academia. A grade of 5 for a category corresponds to a nigh-perfect asset in that respect (e.g., ``overqualified" ramp rate), while a grade of 1 corresponds to an effectively unreliable asset (e.g., nigh-negligible ramp rate). 

\begin{table}[b!]
\captionsetup{font=small}
\large
 \centering
\captionof{table}{Flexibility overview of energy assets. Each flexibility category includes the specific services as shown in Fig. 2}
\scalebox{0.58}{
\begin{tabular}{|c|c|c|c|c|}
\midrule
\textbf{Technology} & \textbf{Ramp} & \textbf{Dynamic} & \textbf{Steady-state} & \textbf{Empirical}\\
 & \textbf{capabilities} & \textbf{capabilities} & \textbf{capabilities} & \textbf{standing}\\
\midrule
Batteries & 4 & 5 & 2 & 3.7 $\pm$ 1.2 \\
\hline
CCGT & 2 & 1 & 1 & 2 $\pm$ 0.8 \\
\hline
OCGT & 3 & 2 & 4 & 3 $\pm$ 0.8 \\
\hline
Solar & 3 & 3 & 2 & 2.7 $\pm$ 0.5 \\
\hline
Wind & 3 & 3 & 3 & 3 $\pm$ 0 \\
\hline
Hydro & 3 & 3 & 4 & 3.3 $\pm$ 0.5 \\
\hline
Pump storage & 3 & 4 & 4 & 3.7 $\pm$ 0.5 \\
\hline
Nuclear & 1 & 1 & 4 & 2 $\pm$ 0.9 \\
\hline
Biomass & 1 & 1 & 2 & 1.3 $\pm$ 0.5 \\
\hline
Demand response & 4 & 4 & 3 & 3.7 $\pm$ 0.5 \\
\hline
\bottomrule
\end{tabular}
}
\label{RobustTable}
\end{table}

\begin{table}[b!]
\captionsetup{font=small}
\large
 \centering
\captionof{table}{Security overview of energy assets. Each security category includes the specific services as shown in Fig. 2.}
\scalebox{0.58}{
\begin{tabular}{|c|c|c|c|c|}
\midrule
\textbf{Technology} & \textbf{Ramp} & \textbf{Dynamic} & \textbf{Steady-state} & \textbf{Empirical}\\
 & \textbf{capabilities} & \textbf{capabilities} & \textbf{capabilities} & \textbf{standing}\\
\midrule
Batteries & 3 & 2 & 5 & 3.3 $\pm$ 1.2 \\
\hline
CCGT & 5 & 5 & 3 & 4.3 $\pm$ 0.9 \\
\hline
OCGT & 4 & 5 & 4 & 4.3 $\pm$ 0.5 \\
\hline
Solar & 3 & 3 & 4 & 3.3 $\pm$ 0.5 \\
\hline
Wind & 4 & 4 & 5 & 4.3 $\pm$ 0.5 \\
\hline
Hydro & 4 & 4 & 3 & 3.7 $\pm$ 0.5 \\
\hline
Pump storage & 5 & 4 & 4 & 4.3 $\pm$ 0.5 \\
\hline
Nuclear & 3 & 5 & 5 & 4.3 $\pm$ 0.8 \\
\hline
Biomass & 2 & 3 & 2 & 2.3 $\pm$ 0.5 \\
\hline
Demand response & 2 & 3 & 4 & 3 $\pm$ 0.8 \\
\hline
\bottomrule
\end{tabular}
}
\label{RobustTable1}
\end{table}

Let us construct a simple example based on the above. Imagine a random power producing installation of unknown device composition and technical characteristics. At every market-clearing time-step, the owner of the installation submits a unified power-price bid to the ``flexicurity market" with the following characteristics:

\begin{itemize}
    \item Security aspect (Quant-Price)$^{security}$: X MW that can contribute to steady-state security (adequacy of supply, stable import/export limits), Y MW that can contribute to dynamic security (fluctuating import/export limits) and Z MW that can contribute to managing the grid’s security parameters (active power quantities, reactive capabilities, inertia-related products). A key difference from the current market setting is that \textit{quantities X, Y, Z may overlap}. Based on how feasible this overlap is (this depends on the technical characteristics of the energy asset), the offer may range from e.g., ``I can provide \textit{Y} for a specific services", to ``I can provide \textit{f(X,Y,Z)} for any security-related service and \textit{X + Y + Z – f(X,Y,Z)} for specific services, and finally all the way up to ``I can provide \textit{X + Y + Z} for any security-related service that may be required". 
    \item Flexibility aspect (Quant-Price)$^{flexibility}$: A MW that can contribute to steady-state security (notice to deliver, maximum deliveries), B MW that can contribute to dynamic flexibility (notices to deviate, minimum zero/non-zero times) and C MW that can contribute to managing the grid’s ramping requirements. Similarly to the security aspect, based on how feasible the overlap between services is the offer may again range from e.g., ``I can provide \textit{B} for a specific services", to ``I can provide \textit{g(A,B,C)} for any flexibility-related service and \textit{A + B + C – g(A,B,C)} for specific services, and finally all the way up to ``I can provide \textit{A + B + C} for any flexibility-related service that may be required"
    \item Final bid (Quant-Price)$^{final}$: Based on how feasible the overlap between flexibility and security is, the final offer effectively revolves around ``I can provide \textit{h\{(Quant-Price)$^{security}$, (Quant-Price)$^{flexibility}$\}=F} for any security-related or flexibility-related service that may be required, and \textit{k(X + Y + Z + A + B + C) - F} for specific services". 
\end{itemize}

So, what does the above example show? For starts, it shows that it is possible to open the door for market players to participate in multiple grid-related markets at the same time. For instance, instead of a conventional plant being restricted to maintain a (almost always excessive amount of) spinning reserve exclusively for secondary frequency control, the plant could under the proposed framework participate in flexibility-related sub-markets or even participate in traditional buy-sell schemes, while at the same time maintaining a smaller yet reliable regulatory presence in the system. This would effectively translate to increased levels of competition across multiple levels, and reduced costs for grid operators from not treating every aspect of their job as a separate entity. The example also shows that if we choose to bundle all services to a common product, flexibility and security could, in theory, be evaluated in shorter time horizons, thus moving closer to the financial markets’ paradigm, i.e., constant, instantaneous real-time transactions. This has been the wish for market players for quite some time.  

A massively important aspect of this approach is that it is effectively technology-agnostic, a major goal for system operators and free-market advocates. This kind of approach arguably brings about the fairest way of operating markets and, while substantially elevating efficiency; technology-agnostic approaches have in fact already been or are being implemented in many markets, e.g., Belgium and the U.K. 

\section{Towards the 2050 smart grid}

The authors would like to close their work by contextualizing the flexicurity-based framework within the 2050 strategy. The 2050 strategy has unquestionably become one of the most, if not the most, crucial drivers of power systems research and power grid-based industrial developments.  It is thus vital to show how a straightforward transition to the currently envisioned power grid could turn out, under today’s plan and market practices. In addition, to show how a flexicurity-based market could achieve monumental shifts in how this transition can be achieved, an alternative strategy to reach 2050-akin goals is also presented. In the authors’ opinion, the proposed strategy is in many respects a superior alternative, at least on a conceptual basis.    

\begin{figure}[b!] 
	\centering
	\scalebox{0.4}{\includegraphics[]{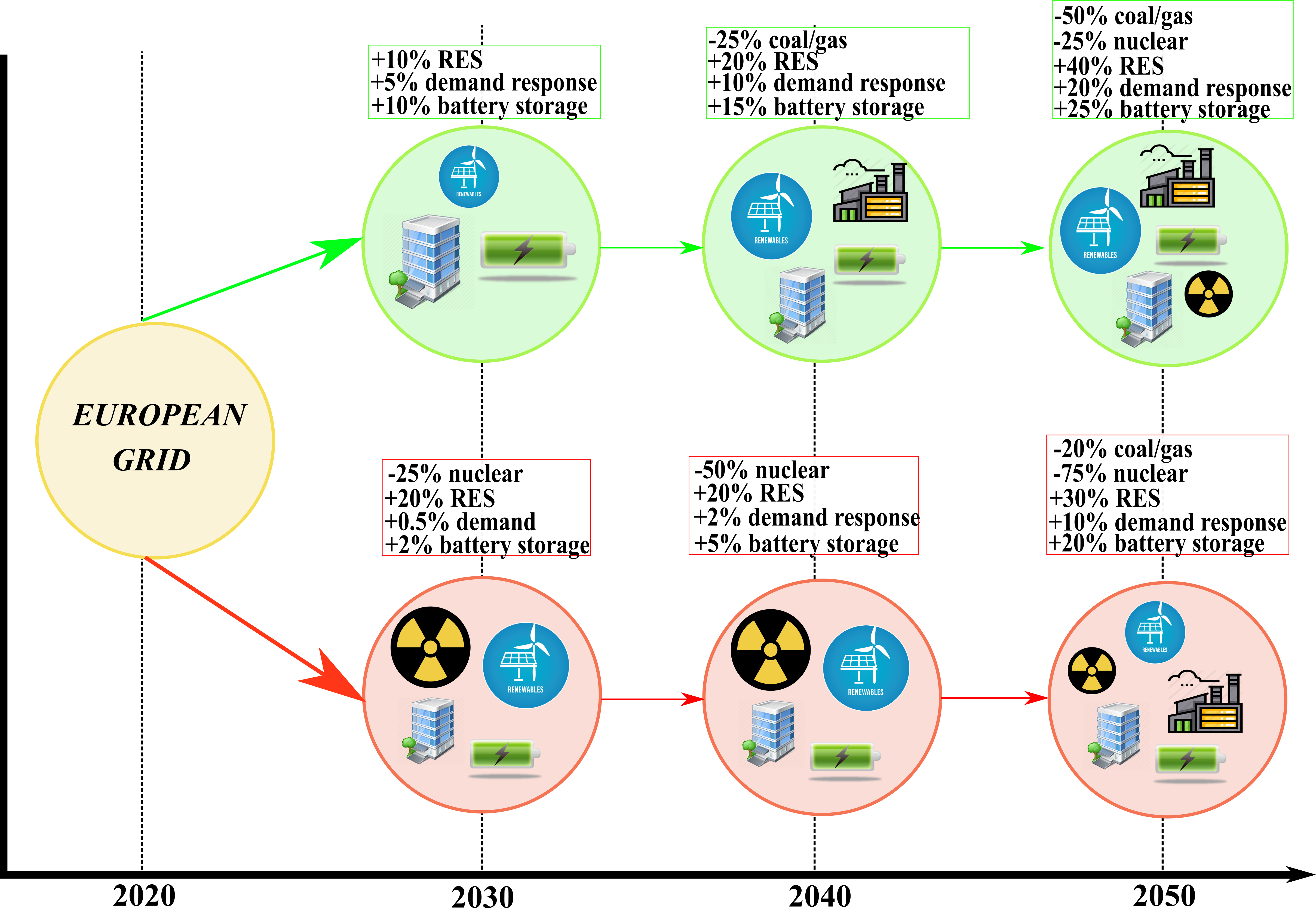}}
	\caption{Two different strategies for reaching the 2050 goals: in red, a steady and aggressive transformation based on conventional market practices. In green, an initially cautious transformation that later invests heavily on the so-called ``game changers" (battery storage, demand response). This strategy is based on the envisioned flexicurity-based market practices.}
	\label{Techniques2}
\end{figure}

We now refer the reader’s attention to Fig. 3, which presents two different strategies for transitioning to the envisioned power grid within the next 3 decades. We begin with the strategy denoted in red. This is a relatively aggressive strategy (in line with what EU directives delineate), which involves the consistent decommissioning of nuclear and coal/gas plants, the steady investment in RES and a slow, almost exclusively driven by private investments, developments of ``game-changers", i.e., battery storage and demand response.  The final state of the power grid would be one with huge penetrations of RES, relatively few nuclear plants, a reduced force of goal/gas plants and some batteries and flexible demand to support the system. While said final state is desirable from a composition perspective (plus, the system’s flexibility does also increase), the security of the system is consistently worsened. The steady removal of power plants contributes to the development of a more ``fragile" system, meaning operators will become even more zealous in their efforts to arm it against extreme events. The over-conservative mentality has no reason to be lifted; in fact, we may even observe coordinated efforts to drastically reduce the degrees of freedom in the grid, i.e., eliminate all possible variability by limiting the range of control variables. Unless the forecasting tools for RES are of exceptional quality, the resulting power system will either consistently find itself in perilous positions or will ``lock out" far more market participants (excessive procurement of reserves) to avoid problematic conditions. Of course, neither situation is viable, for either financial or technical reasons. 

Contrariwise,  the more cautious flexibility-based strategy calls for a coordinated, co-development of security and flexibility. Aside from the reserved increase in RES penetration, the first major changes would include massive investments in the so-called ``game-changers", which would provide huge amounts of potential flexibility, while at the same time relieving the system of some of its stress, automatically making it more secure. Having established a more solid power grid, the rollout of RES could be racked up safely, while some initial decommissioning of conventional plants could begin. The presence of ``game-changers" would naturally have to be consistently strengthened, to avoid the pitfall of reverting to a previous, less desirable state.  Finally, after having established a reliably ``flexicure" system (massive penetration of ``game-changers", strong regulating presence of nuclear plants), the coordinated decommissioning of conventional plants could begin without jeopardizing the power grid. By that point, the flexicurity market would be mature and interconnected enough to solidly stand and support power systems with contribution from energy assets of every conceivable nature and size.

While no definitive generalizable conclusions can safely be made yet, the one constant is that the current market framework was not designed to accommodate the 2050 vision. One could not be blamed for thinking that if said framework does not drastically adapt to its new needs, the 2050 strategy could sadly stay only on paper. On the other hand, a flexicurity-based transition would effectively call for more cautious planning, less disruptive transitions and would be much more positive towards the adoption of technologies with huge potential for contributing decisively to grid management.

\section{Conclusions and next steps}

This work made a first empirical attempt to showcase the main inefficiencies that are currently encountered in power markets. The attempt was almost exclusively qualitative, and yet of no less importance than others; the current status-quo is clearly problematic, or at the very least not as efficient as it could be. The call for a new business model appears to be more urgent now that ever.  The authors, perhaps even crudely or inelegantly, proposed using a unified ``flexicurity" product for securing and managing the power grid, and offered some ideas on how this system could be implemented and what its implications would be for the 2050 strategy. These ideas are still immature and require proper investigation before making any transition to a new market paradigm. Nonetheless, collecting the concerns of hundreds of people from academia and industry, and showcasing them to the community was, in our humble opinion, long overdue.

Having made a first crude attempt to communicate the need for a flexicurity-based power market, the concept must now be further refined and ``mathematized". Quantifying the characteristics of flexicurity in a concrete way and performing a proof-of-concept comparison with the current market framework and practices is the logical next step. This will pave the way for developing sophisticated flexicurity-based optimization models, opening new research avenues. It is the authors’ intent to pursue this noble goal; in case they fall short of succeeding in this undertaking, the hope is that researchers will step in to tackle this new research challenge.

\bibliographystyle{unsrt}

\end{document}